\documentclass[10pt,technote]{IEEEtran}

\usepackage{cite}
\usepackage{amsmath,amssymb,amsfonts}
\usepackage{algorithmic}
\usepackage{graphicx}
\usepackage{textcomp}
\usepackage{siunitx}
\usepackage{lipsum}
\usepackage{tikz}
\usepackage[nolist,nohyperlinks]{acronym}
\usepackage{subfigure}
\usepackage{multirow}
\usepackage{stfloats}
\usepackage{float}

\newcommand{\red}[1]{\textcolor{red}{#1}}
\definecolor{darkgreen}{RGB}{0,204,56}
\newcommand{\green}[1]{\textcolor{darkgreen}{#1}}
\definecolor{bluerevision}{RGB}{0,0,0}
\newcommand{\blue}[1]{\textcolor{bluerevision}{#1}}

\def\BibTeX{{\rm B\kern-.05em{\sc i\kern-.025em b}\kern-.08em
    T\kern-.1667em\lower.7ex\hbox{E}\kern-.125emX}}
\begin{document}

\begin{acronym}
    \acro{SGH}{Standard gain horn}
    \acro{LoS}{Line-of-Sight}
    \acro{FFC}{face-forward crossing}
    \acro{CFC}{cross-forward crossing}
    \acro{UTD}{uniform theory of diffraction}
    \acro{FI}{Floating-intercept}
    \acro{CI}{Close-in}
    \acro{CIF}{\ac{CI} with a frequency-dependent term}
    \acro{ABG}{Alpha-beta-gamma}
    \acro{RV}{random variable}
    \acro{DKED}{double knife-edge diffraction}
    \acro{IFBW}{intermediate frequency bandwidth}
    \acro{3GPP}{3rd Generation Partnership Project}
    \acro{DTMKE}{double-truncated multiple knife-edge}
\end{acronym}

\title{Experimental Assessment of Human Blockage at sub-THz and mmWave Frequency Bands}

\markboth{\footnotesize This work has been accepted for publication in \textit{IEEE Transactions on Vehicular Technology}. DOI: TVT.2026.3682016}%
{\footnotesize This work has been accepted for publication in \textit{IEEE Transactions on Vehicular Technology}. DOI: TVT.2026.3682016}

\author{Juan E. Galeote-Cazorla, Alejandro Ramírez-Arroyo, Jose-Maria Molina-Garcia-Pardo, Maria-Teresa Martinez-Ingles and Juan F. Valenzuela-Valdés
\thanks{\noindent 
This work has been supported by grant PID2024.157242OB.C44 funded by MCIN/AEI/10.13039/501100011033 and by ERDF/EU. It has also been supported by grants PID2020-112545RBC54, PID2022-136869NB-C32, and PDC2023-145862-I00, funded by MCIN/AEI/10.13039/501100011033 and by ERDF/EU and the European Union NextGenerationEU/PRTR and grant DGP\_PIDI\_2024\_00736 by Junta de Andalucía; and in part by the predoctoral grant FPU22/03392 (\textit{Corresponding author: Juan E. Galeote-Cazorla}).}
\thanks{Juan E. Galeote-Cazorla and Juan F. Valenzuela-Valdés are with the Department of Signal Theory, Telematics and Communications, Research Centre for Information and Communication Technologies (CITIC-UGR), Universidad de Granada, 18071 Granada, Spain (e-mail: juane@ugr.es; juanvalenzuela@ugr.es).}
\thanks{Alejandro Ramírez-Arroyo is with the Department of Electronic Systems, Aalborg University (AAU), 9220 Aalborg, Denmark (e-mail: araar@es.aau.dk).}
\thanks{Jose-Maria Molina-Garcia-Pardo is with the Information Technologies and Communications Department, Universidad Politécnica de Cartagena, 30202 Cartagena, Spain (e-mail: josemaria.molina@upct.es).}
 \thanks{Maria-Teresa Martinez-Ingles is with the  Departamento de Automática, Ingeniería Eléctrica y Tecnología Electrónica, ETS de Ingeniería Industrial, Universidad Politécnica de Cartagena, 30202 Cartagena, Spain (e-mail: mteresa.martinez@upct.es).}
 }

\maketitle

\begin{abstract}
    The fifth generation (5G) of mobile communications relies on extremely high data transmission rates utilizing a wide range of frequency bands, including FR1 (sub-6\:GHz) and FR2~(mmWave). Future mobile communications systems are envisaged to operate at the electromagnetic spectrum beyond FR2, above 100\:GHz, known as sub-THz band. These new frequencies open up challenging scenarios where communications \blue{will have to} rely on a major contribution such as the line-of-sight~(LoS) component. To the best of the authors' knowledge, for the first time in \blue{the} literature this work studies the human blockage effects over an extremely wide frequency band from 75\:GHz to 215\:GHz considering: (i) the distance between the blocker and the antennas and (ii) the body size and orientation. The obtained results are fitted to modifications of the classical path loss models and compared to 3GPP alternatives. The average attenuation increases from 42\:dB to~56\:dB when frequency rises from 75\:GHz to 215\:GHz. \blue{On the other hand,} an 18\:dB increment in the received power is observed when the Tx--Rx separation is increased from 1\:m to 2.5\:m. Finally, variations of up to 4.6\:dB are found depending on the blocker's orientation.
\end{abstract}

\begin{IEEEkeywords}
    sub-THz, mmWave, human blockage, channel modeling, diffraction, blockage gain, propagation
\end{IEEEkeywords}

\section{Introduction}\label{sec:introduction}
\IEEEPARstart{T}{he} evolution of wireless communication systems is evolving toward higher data rates, system capacities, carrier frequencies, and bandwidths \cite{Rikkinen2020}, \cite{Han2022}. The current fifth-generation (5G) and the future sixth-generation (6G) of wireless systems consider not only millimeter-waves (mmWave) frequencies but also \linebreak sub-THz (100\:GHz -- 300\:GHz) and THz (0.1\:THz -- 10\:THz) bands as key enablers for meeting their strict requirements~\cite{Dang2020}.

Systems operating at low frequencies (sub-6\:GHz) can rely on the main propagation mechanisms such as reflection, diffraction, and scattering \cite{Andersen1995}. In contrast, this is not the case for sub-THz and THz bands. At these frequencies, the electrical dimensions of the objects increase significantly, leading to a detriment of diffraction effects. Moreover, diffuse scattering becomes more relevant due to the roughness of the surfaces. Together, these characteristics result in weaker specular components compared to lower frequencies \cite{Piesiewicz2007, THz_material}. These behaviors, in addition to the reduced aperture size of antennas in the THz region, imply that communications must rely on the line-of-sight (LoS) component~\cite{Guan2021}. In mobile communication scenarios, this contribution can occasionally be shadowed. Since the radius of the first Fresnel zone is proportional to the wavelength, the human body can easily block the LoS path.

As a consequence of the aforementioned reasons, there is a growing interest in assessing the impact of human blockage on the performance of communication systems. Thus, accurate deterministic channel models for human body shadowing are needed. Some examples of these models have been proposed at mmWave and THz bands in the state-of-the-art. In \cite{Karadimas2013}, the variations in received power due to human activity were statistically characterized in the time domain for 60\:GHz indoor short-range wireless links. In~\cite{Dalveren2021}, the authors present a simple approach to characterize the scattering effects of nearby human bodies on a short-range indoor link at 28\:GHz, while the link is completely blocked by another body. In~\cite{Virk2020}, human blockage measurements were conducted in an anechoic chamber at 15\:GHz, 28\:GHz, and 60\:GHz frequencies employing 15 human subjects of different sizes and weights. Additionally, an effective 3-D human blockage model based on a \ac{DTMKE} \linebreak scheme was proposed. 

Indeed, diffraction has been used in many studies to model human shadowing. Knife-edge diffraction (KED) models are commonly employed in literature \cite{ITU2019}, \cite{MacCartney2016}. However, they have the drawback that the 3D models are limited to a geometry based on metallic planes. Furthermore, the scattering contribution from the human body cannot be separated from the LoS path. The \ac{UTD} offers moderate computational cost and allows flexible 3D modeling with elements such as screens, \blue{hexagonal} cylinders, and circular cylinders \cite{Pathak1980}. In contrast, physical optics~(PO) approximation provides high accuracy when modeling the most detailed circumference of the cross-section (CCS) as a human phantom, but its computational cost is unacceptable for dynamic scenarios. Most of the existing studies focus on mmWave, while a few deal with frequencies at sub-THz bands. For instance, in \cite{Zhang2023, Takagi2022} the authors present results evaluating reflection and blockage losses caused by the human body in the 140\:GHz and 300\:GHz~bands, respectively.

In this work, a study on human blockage of the LoS is performed over a wide frequency range from 75\:GHz to 215\:GHz including both mmWave and sub-THz bands. After a thorough analysis in terms of frequency, distance between the antennas, blocker orientation, and human body dimensions, the blockage gain parameter is proposed to quantify the impact on the LoS signal\blue{.} This parameter is defined as the maximum decrease \blue{in} the signal power due to a blockage event. Then, the blockage gain is modeled applying a modified version of classical path loss models as {a} function of frequency and distance, offering an alternative to the geometry-based approaches. Accurate results are obtained, with reduced standard deviation and high channel predictability. Finally, our proposals are compared with the geometrical models defined by the 3rd Generation Partnership Project~(3GPP) known as 3GPP-A and \linebreak 3GPP-B~\cite{3GPP2022}, that have higher complexity and yield less accurate predictions. Quantifying blockage effects in the sub-THz and THz bands is essential for the development of personal area networks~(PANs) and body area networks~(BANs) systems in future cellular generations. The blockage gain parameter is conceived as a figure of merit to easily evaluate a link-budget considering human blockage effects as an additional contribution to the path loss and other attenuation sources.

The remainder of the article is organized as follows. Section~II describes the measurements acquisition process and the subsequent post-processing. Section III analyzes the influence of the frequency, distance between the antennas, blocker orientation and human body dimensions. Then, it presents the blockage gain models and their fit with the experimental results followed by a comparison with other models from the literature. Finally, Section~IV provides the conclusions \linebreak of this work.

\section{Measurements Acquisition and Processing}
\begin{figure}[t]
    \centering
    \includegraphics{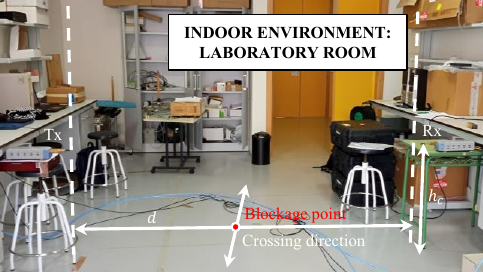}
    \caption{Panoramic \blue{view} of the measurement setup at the UPCT.}
    \label{fig:setup_image}
\end{figure}

Measurements have been carried out in a laboratory at the Universidad Politécnica de Cartagena~(UPCT), with dimensions of 8\:$\times$\:4.8\:$\times$\:3.5\:m\textsuperscript{3}. The environment is furnished with desks, chairs, shelves and closets, and it is equipped with several electronic devices such as computers and measurement instruments (see Fig. \ref{fig:setup_image}). Transmitter (Tx) and receiver~(Rx) are aligned both horizontally and vertically, separated by a distance $d$\:$\in$\:\{1, 1.75, 2.5\}\:m and placed at a height $h_c$\:=\:1\:m above the ground\footnote{Preliminary experimental tests suggest that the influence of antenna height is significantly lower compared to other factors such as frequency or the distance between the antennas.}, which is close to the hip height of the human body. This setup resembles a device-to-device~(D2D) communication scenario within an indoor environment, where fading due to the transit of humans is typically observed.

The channel sounder consists in a vector network analyzer (VNA, model R\&S\textsuperscript{\textregistered} ZVA67) and several frequency converters (models R\&S\textsuperscript{\textregistered} ZVA-Z110E, ZC170 and ZC220), enabling signal acquisition in the range from 75\:GHz to 220\:GHz. The Tx and Rx use standard gain horn (SGH) antennas with a gain of 20\:dB and a half-power beamwidth (HPBW) of 18º in both the H and E planes at mid-band (models Flann SGH Series 240 \#27240 and \#29240, for the 75\:GHz--110\:GHz and 110\:GHz--215\:GHz bands respectively). They are placed to operate with vertical polarization for both transmission and reception. The VNA operates in continuous wave (CW) mode, with a sampling frequency of 100\:Hz (corresponding to a time step of 10\:ms). The number of points is set to 2\:048, which yields a maximum excess time of 20.47\:s. Nine different frequencies are evaluated, equispaced by 17.5\:GHz in the range from 75\:GHz to 215\:GHz. The most restrictive dynamic range is above 80\:dB, and it is observed for the highest frequency.

Referring to Fig. \ref{fig:setup_image}, the measurement methodology involves a human subject crossing along the perpendicular direction to the Tx--Rx link at its midpoint. Under these conditions, the blockages are produced in the far-field region of the antennas even for the shortest distance of 1\:m. Additionally, the distances are large enough to assume that the variation of the antennas HPBW with frequency have no significant impact on the measurements. Two subjects participated in the experiment, performing 2 head-on and 2 sideways crossings per frequency and distance. Therefore, the total number of captured blockage events is 2\:$\times$\:4\:$\times$\:9\:$\times$\:3\:=\:216. In all cases, the crossing speed was constant and took a value of 45\:cm/s (1.62\:km/h), lower than the typical values for pedestrian speed\footnote{The typical pedestrian speed varies between 4\:km/h and 6\:km/h \cite{Murtagh2020}.} to accentuate the blockage effects. The first subject is 1.76\:m tall, weighs 74\:kg and has a hip width of 42\:cm, while the second subject is 1.74\:m tall, weighs 91\:kg and has a hip width of 44\:cm. This diversity, along with the different crossings, frequencies, and distances, allows for the generation of several attenuation patterns.

\begin{figure}[t]
    \centering
    \includegraphics[width=0.8\linewidth]{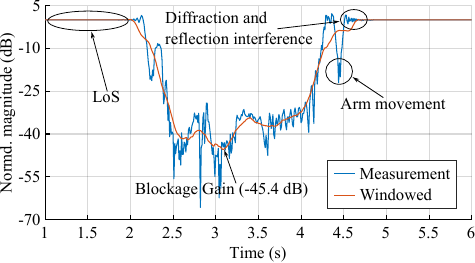}
    \caption{Example of blockage event measured for $f$\:=\:75\:GHz and $d$\:=\:1\:m (normalized with respect to the LoS component). The red line results from removing the fast fading effects of the measurement.}
    \label{fig:blockage_example}
\end{figure}

An example of a blockage event trace is presented in Fig.~\ref{fig:blockage_example}, captured at a frequency $f$\:=\:75\:GHz, a distance $d$\:=\:1\:m and with a sideways crossing. The signal magnitude is normalized to the power of the LoS component (including antenna gains), resulting in the continuous 0\:dB value observed at the beginning and end of the trace. In the central portion, the LoS component is fully blocked, which implies a degradation of the received power and increased fast fading effects. Additionally, several phenomena such as arm movements (which cause a partial signal recovery) as well as diffraction plus reflection interferences are also observed. In order to characterize the blockage, fast fading effects have to be removed. This can be done by averaging over the local mean power with a sliding window of 16 samples\footnote{Given a crossing speed of 45\:cm/s, this window size is equivalent to 32$\lambda$ at the midband frequency. This value is within the range from 20$\lambda$ to 40$\lambda$, which retains the trace slow-fading information \cite{Lee1985}.} \cite{Lee1985}. Then, the concept of blockage gain (BG) naturally arises. We define it as the minimum gain observed along the blockage after removing the fast fading effects. In the example shown in Fig. \ref{fig:blockage_example}, the blockage gain takes a value of $-$45.5\:dB.

\begin{figure}[t]
    \centering
    \subfigure[]{\includegraphics[width=0.8\linewidth]{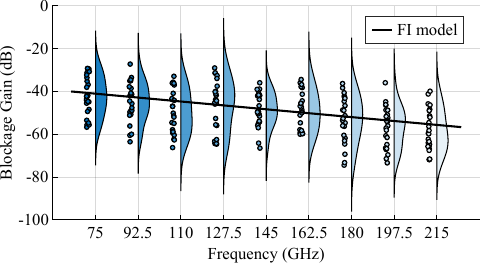}}
    \subfigure[]{\includegraphics[width=0.8\linewidth]{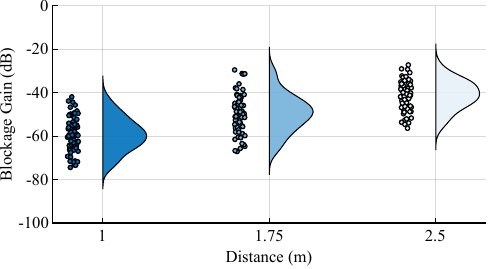}}
    \caption{BG samples (scatter plot) and probability density estimate (violin diagram) per studied (a) frequency and (b) distance. The fit with the modified floating-intercept (FI) model is included as a black line in subfigure (a).}
    \label{fig:boxplot_freq}
\end{figure}

\section{Analysis and Modeling of Blockage Gain}
Applying the aforementioned definition, the BG parameter was computed for each captured blockage event. Next, it was analyzed in terms of the factors considered in our experiment. As shown in Fig. \ref{fig:boxplot_freq}(a), the BG exhibits a clear inversely proportional relation with frequency. In particular, median values of $-$43.0\:dB and $-$59.7\:dB are obtained at the lowest and highest frequencies respectively (75\:GHz and 215\:GHz). While the free-space signal attenuation along this frequency band ranges \blue{by} 9.1\:dB, additional losses of up to 16.7\:dB are observed during the blockage events. This is due to the growing electrical dimensions of the blocker at higher frequencies, which increases the signal attenuation. Note that the high variability of the blockage gain between samples given a certain frequency in Fig. \ref{fig:boxplot_freq}(a) is due to taking into consideration the other variables of the measurement campaign, i.e., (i) distance between antennas, (ii) body orientation, and (iii) human subject. Additionally, the figure represents with a dark line the fit between the measurements and the modified floating-intercept~(FI) model discussed later in Section III.B.

An analogous analysis was performed for the remaining factors. Particularly, Fig. \ref{fig:boxplot_freq}(b) presents the blockage gain samples and distribution obtained for all frequencies and blockage events for each Tx-Rx distance. In contrast to the frequency trend, the blockage gain \blue{decreases} as the distance between the antennas becomes shorter. For each configuration (1\:m, 1.75\:m, and 2.5\:m), the median BG values are $-$59.5\:dB, $-$49.4\:dB, and $-$41.2\:dB respectively; representing an 18.3\:dB difference from $d$\:=\:1\:m to $d$\:=\:2.5\:m. In this case, the antennas aperture plays a fundamental role. As the distance increases, the relative dimensions of the blocker with respect to the antenna beam decreases, enhancing the diffraction around the edges of the human~body. On the other hand, in terms of the human subject, the discrepancy between the median BG values is 7.2\:dB, indicating clear differences in their physical constitutions. Finally, regarding the body orientation, median values of $-$51.5\:dB and $-$46.9\:dB are obtained for head-on and sideways crossings respectively, yielding a difference of 4.6\:dB. This result highlights the impact of the human body dimensions along the LoS direction, which is higher in the head-on case than in the sideways case. Finally, an analysis of variance (ANOVA) has been performed to quantify the impact of the considered factors on the BG parameter. From this, $p$-values lower than 0.01 were obtained for all \blue{factors} \linebreak \blue{---}frequency, distance between the antennas, human subject and blocker orientation\blue{---} which implies that these are significant at the 99\%.

\subsection{Blockage Gain models}
In the state-of-the-art, there are numerous proposals for modeling path loss as a simple function of frequency and distance. However, this is not the case for the blockage losses, which are usually described by diffraction-based models as the~\ac{DTMKE} \cite{Virk2020}. These models are highly dependent on the geometrical description of the analyzed scenario, which translates into an increased complexity. 

As mentioned in previous paragraphs, blockage gain exhibits a clear dependence on frequency and distance. These numerical factors can be incorporated into regression models, allowing us to describe the blockage gain across the wide range of studied frequencies on the mmWave and sub-THz bands. Seeking a simple description of the blockage behavior, we propose modeling the BG using modified versions of the classical path loss models \cite{Rappaport2017}:
\begin{itemize}
    \item \textit{\ac{FI}.} Blockage gain depends logarithmically on frequency, and it is given by
    \begin{equation}\label{eq:human_blockage_fi}
        \mathrm{BG}(f) = A + 10n\log_{10}\left(\frac{f}{1\:\mathrm{GHz}}\right) + \chi_\sigma,
    \end{equation}
    where $A$ is the intercept\footnote{The intercept parameter represents the blockage gain value returned by the model at the frequency and distance of reference. In this case, for $f$\:=\:1\:GHz and for a distance $d$\:=\:0. It can be interpreted as a bias or offset in the model.}, and $n$ is the loss exponent.
    
    \item \textit{\ac{CI}.} Logarithmic and linear dependencies with frequency and distance, respectively, are assumed. Then,
    \begin{equation}\label{eq:human_blockage_ci}
        \mathrm{BG}(d,f) = \phi d + 10m\log_{10}\left(\frac{f}{1\:\mathrm{GHz}}\right) + \chi_\sigma,
    \end{equation}
    where $\phi$ models the distance-dependent term, and $m$ represents the frequency loss exponent.

    \item \textit{\ac{ABG}.} This model is analogous to the \ac{CI} model but introduces an intercept $\beta$. That is,
    \begin{equation}\label{eq:human_blockage_abg}
        \mathrm{BG}(d,f) = \alpha d + \beta + 10\gamma\log_{10}\left(\frac{f}{1\:\mathrm{GHz}}\right) + \chi_\sigma,
    \end{equation}
    where $\alpha$ is the distance gain factor, and $\gamma$ is the frequency loss exponent.
    
    \item \textit{\ac{CIF}.} This model introduces a cross frequency-distance dependence. Then,
    \begin{multline}\label{eq:human_blockage_cif}
        \mathrm{BG}(d,f) = a\left(1 + b\:\frac{f - f_0}{f_0}\right)d \\ + 10c\log_{10}\left(\frac{f}{1\:\mathrm{GHz}}\right) + \chi_\sigma,
    \end{multline}
    where $a$ models the distance-dependent term, $b$ is the cross frequency-distance dependence factor, and $c$ is the frequency loss exponent. Additionally, $f_0$\:=\:145\:GHz is the central frequency of the studied band.
\end{itemize}
All the models expressions are in logarithmic scale (i.e., units of decibels) and include a stochastic term $\chi_\sigma$ to capture factors not explicitly modeled such as blocker orientation, human subject differences, and other masked influences. It is assumed to be a zero-mean Gaussian distributed \ac{RV} with variance~$\sigma^2$. The main differences compared to the classical path-loss models are (i) substitution of logarithmic dependence on distance by a linear one, and (ii) replacing the free-space path loss reference by a one-freedom degree power law in frequency. These changes allow us to describe the blockage effects in a straightforward manner using a reduced number of parameters. Finally, these models estimates BG for a blocker located at the midpoint of the link. Nevertheless, for locations closer to the Tx/Rx antenna, they return a lower bound of BG by considering an equivalent distance $d$ equal to twice the distance from the blocker to the closest antenna.

In order to compare our proposed models with geometry-based alternatives, we also apply the 3GPP blockage models 3GPP-A and 3GPP-B \cite{3GPP2022}. Although these are defined for frequencies up to 100\:GHz, they can be analytically extended to the upper bound of our measurement range, i.e., up to 215\:GHz. The blockage gain estimation is based on modeling diffraction around metallic planes placed in a 3D space. The 3GPP-A model adopts a stochastic approach that requires only the angles formed by the Rx and the edges of the planes. In contrast, the 3GPP-B model follows a deterministic approach that needs all distances between the edges of the metallic planes and both Tx and Rx.

\begin{figure}[t]
    \centering
    \usetikzlibrary{arrows.meta}
        \begin{tikzpicture}[scale=0.85]
            \node[] at (2.6125,0) {\footnotesize \textbf{Front view}};
            \draw (-3.5,-2.225) -- (3.5,-2.225);
            
            \draw (0,-0.125) circle (0.2125);
            \draw (-0.2,-0.4) -- (0.2,-0.4) arc (90:0:0.3) -- (0.5,-1.25) arc (360:180:0.1);
            \draw (-0.2,-0.4) arc (90:180:0.3) -- (-0.5,-1.25) arc (180:360:0.1);
            \draw (0.3,-1.25) -- (0.3,-0.75) arc (0:180:0.025) -- (0.25,-2.125) arc (360:180:0.1) -- ++(0,0.75) arc (0:180:0.05);
            \draw (-0.3,-1.25) -- (-0.3,-0.75) arc (180:0:0.025) -- (-0.25,-2.125) arc (180:360:0.1) -- ++(0,0.75);
            
            \draw (-1.5,-1.25) rectangle (-2.75,-2.225) (1.5,-1.25) rectangle (2.75,-2.225);
            \draw (-1.5,-1.25) rectangle (-2,-1) node[above] {\footnotesize Tx} (1.5,-1.25) rectangle (2,-1) node[above] {\footnotesize Rx};
            \draw (-1.5,-1.12) -- (-1.45,-1.12) -- (-1.4,-1.08) -- (-1.4,-1.16) -- (-1.45,-1.13) -- (-1.5,-1.13);
            \draw (1.5,-1.12) -- (1.45,-1.12) -- (1.4,-1.08) -- (1.4,-1.16) -- (1.45,-1.13) -- (1.5,-1.13);
            
            \draw[densely dashed] (-1.4,-1.2) -- (-1.4,-3) (1.4,-1.2) -- (1.4,-3);
            \draw[{Latex[scale=0.5]}-{Latex[scale=0.5]}] (-1.4,-2.9) -- (1.4,-2.9) node[pos=0.5,below] {\footnotesize $d$};
    
            \draw[densely dashed] (-3,0.0875) -- (0.6,0.0875);
            \draw[{Latex[scale=0.5]}-{Latex[scale=0.5]}] (-2.9,0.0875) -- (-2.9,-2.225) node[pos=0.5,left] {\footnotesize $h$};
    
            \draw[densely dashed] (-0.5,-1.3) -- (-0.5,-2.5) (0.5,-1.3) -- (0.5,-2.5);
            \draw[{Latex[scale=0.5]}-{Latex[scale=0.5]}] (-0.5,-2.4) -- (0.5,-2.4) node[pos=0.5,below] {\footnotesize $r$};
    
            \draw[densely dashed] (2,-1.125) -- (3,-1.125);
            \draw[{Latex[scale=0.5]}-{Latex[scale=0.5]}] (2.9,-1.125) -- (2.9,-2.225) node[pos=0.5,right] {\footnotesize $h_c$};
    
            \draw[blue,thick,dashed] (-0.5,-2.225) -- (-1.4,-1.125) -- (-0.5,0.0875) -- (1.4,-1.125) -- (-0.5,-2.225);
            \draw[blue,thick] (-0.5,0.0875) -- (-0.5,-2.225);
            \draw[red,thick,dashed] (0.5,-2.225) -- (-1.4,-1.125) -- (0.5,0.0875) -- (1.4,-1.125) -- (0.5,-2.225);
            \draw[red,thick] (0.5,0.0875) -- (0.5,-2.225);
        \end{tikzpicture}
        \begin{tikzpicture}[scale=0.85] 
            \draw[color=white] (-3.5,0) -- (3.5,0);
            \node[] at (2.53,1.125) {\footnotesize \textbf{Top view}};
            \node[] at (-2.25,0) {\footnotesize Tx};
            \node[] at (2.25,0) {\footnotesize Rx};
            \draw (-2.75,-0.75) rectangle (-1.5,0.75) (2.75,-0.75) rectangle (1.5,0.75)
                  (-1.5,0.165) rectangle (-2,-0.165) 
                  (1.5,0.165) rectangle (2,-0.165);
            \draw (-1.5,0.005) -- (-1.45,0.005) -- (-1.4,-0.045) -- (-1.4,0.045) -- (-1.45,-0.005) -- (-1.5,-0.005);
            \draw (1.5,0.005) -- (1.45,0.005) -- (1.4,-0.045) -- (1.4,0.045) -- (1.45,-0.005) -- (1.5,-0.005);
            
            \draw (0,0) ellipse (0.5 and 0.25);
            \draw[densely dashed] (0,0.25) -- (3,0.25) (0,-0.25) -- (3,-0.25);
            \draw[{Latex[scale=0.5]}-{Latex[scale=0.5]}] (2.9,-0.25) -- (2.9,0.25) node[pos=0.5,right] {\footnotesize $w$};
            
            \draw (0,0) circle (0.2125);

            \draw[densely dashed] (-1.4,0) -- (-1.4,-1.5) (1.4,0) -- (1.4,-1.5);
            \draw[{Latex[scale=0.5]}-{Latex[scale=0.5]}] (-1.4,-1.4) -- (1.4,-1.4) node[pos=0.5,below] {\footnotesize $d$};

            \draw[densely dashed] (-0.5,0) -- (-0.5,-1) (0.5,0) -- (0.5,-1);
            \draw[{Latex[scale=0.5]}-{Latex[scale=0.5]}] (-0.5,-0.9) -- (0.5,-0.9) node[pos=0.5,below] {\footnotesize $r$};

            \draw[blue,thick] (-0.5,-0.25) -- (-0.5,0.25);
            \draw[blue,thick,dashed] (-0.5,-0.25) -- (-1.4,0) -- (-0.5,0.25) -- (1.4,0) -- (-0.5,-0.25);
            \draw[red,thick] (0.5,-0.25) -- (0.5,0.25);
            \draw[red,thick,dashed] (0.5,-0.25) -- (-1.4,0) -- (0.5,0.25) -- (1.4,0) -- (0.5,-0.25);
        \end{tikzpicture}
    \caption{Geometrical description of the carried out measurements applied to estimate the blockage gain with the 3GPP models. The metallic planes and diffraction paths are represented by solid and dashed lines respectively.}
    \label{fig:human_blocking_scheme}
\end{figure}

For our specific case, we represent the human blocker as two metallic planes whose height and width are adjusted to match the dimensions of the human body. The blockage gain is then computed as the sum of the individual contributions from each plane. Assuming a typical human, the body dimensions are a height $h$\:=\:1.7\:m, a width $r$\:=\:0.4\:m, and a depth $w$\:=\:0.3\:m. As illustrated in Fig. \ref{fig:human_blocking_scheme}, the blocker is positioned at the midpoint of the Tx--Rx line, where the attenuation is maximum. The metallic planes (shown as red and blue solid lines) are placed on either side of the body, each with dimensions $w$\:$\times$\:$h$. The distance between the antennas is set to the corresponding value of $d$, and both are located at a height $h_c$\:=\:1\:m above the ground. The geometry required to estimate the blockage gain is defined by the diffraction paths indicated by the red and blue dashed lines. For the 3GPP-A model, only the angles formed by these paths at the receiver side are needed. However, the 3GPP-B model requires the full lengths of all the diffraction paths at both transmitter and receiver sides. In both cases, these geometric features are entirely determined by the blocker's dimensions and the Tx-Rx distance.

\subsection{Modeling discussion}
A fit between the models presented in eqs. (1)--(4) and the BG measurements was carried out applying the least squares method. Thus, the optimal parameters are those which minimize the variance $\sigma^2$ of the stochastic term $\chi_\sigma$. 

The FI model is the simplest among those considered, since it only accounts for the operation frequency. Despite the high dispersion in \blue{the} measurements, it successfully captures the observed trend as shown in Fig. \ref{fig:boxplot_freq}. An intercept $A$\:=\:16.0\:dB and a loss exponent $n$\:=\:$-$3.09 were obtained. These results indicate that a frequency increment in the order of 10 leads to a reduction of approximately 30\:dB on the received power. This highlights the significance of human blockage effects at the sub-THz band. At these frequencies, the electrical dimensions of obstacles increase substantially compared to those at sub-6\:GHz and mmWave bands, resulting in higher penetration losses and a reduced impact of diffraction mechanisms. The minimum standard deviation $\sigma$ obtained for the model is $\sigma_{\rm FI}$\:=\:9.7\:dB, which aligns with the aforementioned dispersion seen in Fig. \ref{fig:boxplot_freq}.

\begin{figure*}[t]
    \centering
    \includegraphics[width=0.79\linewidth]{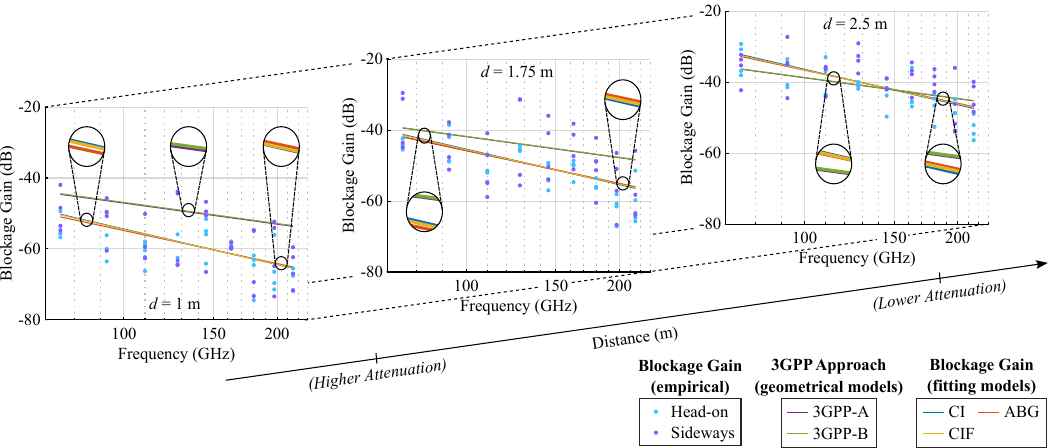}
    \caption{Comparison between the CI, ABG, and CIF blockage gain fitting models (blue, red and yellow lines) with the 3GPP-A and 3GPP-B blockage gain geometry-based models (purple and green lines) for the three studied distances between the antennas. The blue and purple points represent the empirical samples of blockage gain obtained from head-on and sideways crossings respectively. From the zoomed-in views of the curves, it can be observed that the three fitting models closely overlap with each other, as well as the two geometrical models. Additionally, since blockage gain is inversely related to attenuation, the figure clearly illustrates that shorter Tx-Rx distances $d$ are tailored to higher attenuation while longer distances $d$ are associated with lower attenuation.}
    \label{fig:freq_dist_models}
\end{figure*}

The remaining models introduce the distance between the antennas as an additional variable alongside the operation frequency. In the case of the CI model, as with the FI one, only two parameters are involved. After performing the fit, the obtained values were $\phi$\:=\:12.0\:dB/m for the distance gain factor and $m$\:=\:$-$3.32 for the frequency loss exponent. This last parameter is consistent with that obtained for the FI model but is slightly higher due to the absence of an intercept parameter. When the intercept is introduced, the ABG model naturally emerges as a generalization of the CI model. For this model, the obtained parameters were a distance gain factor $\alpha$\:=\:12.1\:dB/m, an intercept $\beta$\:=\:$-$5.2\:dB, and a frequency loss exponent $\gamma$\:=\:$-$3.09. The latter matches the value obtained for the FI model, demonstrating that the ABG model generalizes both the FI and CI models. Lastly, the CIF model modifies the CI model by introducing a cross dependence with frequency and distance. The obtained parameters were a distance gain factor $a$\:=\:12.1\:dB/m, a cross frequency-distance factor $b$\:=\:0.0196, and a frequency loss exponent $c$\:=\:$-$3.32. Since there is no intercept parameter, $c$ is equal to that of the CI model. Regarding parameter~$b$, its small value suggests that distance and frequency are uncorrelated in this scenario, which explains the similarity between the models parameters. Consequently, the obtained $\sigma$ for each fit is practically the same: $\sigma_{\rm CI}$\:=\:6.18\:dB, $\sigma_{\rm ABG}$\:=\:6.17\:dB, and $\sigma_{\rm CIF}$\:=\:6.17\:dB. 

All the aforementioned results are summarized in Fig.~\ref{fig:freq_dist_models}, where the graph of blockage gain versus frequency for each Tx-Rx distance is depicted. The empirical blockage gain samples are represented by blue and purple points, distinguishing between head-on and sideways crossings, respectively. In contrast, the fitted CI, ABG and CIF models correspond with the blue, red and yellow lines. It can be seen that the regression curves almost completely overlap. As \blue{shown in} the zoomed-in regions, slight differences appear in their slopes due to their mathematical expressions and the small differences in the parameter values. Nevertheless, all three models successfully capture the decreasing trend of blockage gain with frequency and the increasing trend with distance. Another important observation is the consistency between the distance gain factors $\phi$, $\alpha$, and $a$ in the CI, ABG and CIF models respectively. They converge to a value \blue{close} to 12\:dB/m, which is explained by the reduced wavelength at the sub-THz bands and the antennas' HPBW. An increase of one meter in distance means separating the antennas by several hundred wavelengths, significantly reducing the relative electrical dimensions of the blocker and thus favoring the diffraction mechanisms around its~edges. 

To evaluate the performance of the models, we have conducted a blind verification with a third human subject who is 1.78\:m tall, weighs 80\:kg and has a hip width of 40\:cm. After post-processing, the root mean squared errors (RMSEs) obtained for the FI, CI, ABG and CIF models were 9.48\:dB, 6.01\:dB, 5.89\:dB and 6.13\:dB respectively. In all cases, these values are lower than the corresponding standard deviations, indicating that the blockage gain measurements for the third person are within the variability range of the models. Therefore, we conclude that their predictions are reliable.

Alternatively to the FI, CI, ABG and CIF models, we have also estimated the blockage gain by applying the 3GPP-A and 3GPP-B models to the geometry presented in Fig.~\ref{fig:human_blocking_scheme}. The obtained RMSEs with respect to the measurements were $\mathrm{RMSE}_{\rm A}$\:=\:8.9\:dB and $\mathrm{RMSE}_{\rm B}$\:=\:9.0\:dB respectively. We have represented the curves in Fig. \ref{fig:freq_dist_models} as the purple and green lines respectively. It can be seen that both models yield similar BG values. For a distance $d$\:=\:1\:m, the 3GPP models underestimate the blockage gain by approximately 10\:dB. This discrepancy diminishes as the distance increases. In the case of $d$\:=\:2.5\:m, these models reasonably approximate the measurements. Therefore, the increase of BG with distance is more gradual in the 3GPP models than in the CI, ABG, and CIF models. This behavior can be explained by considering the Tx and Rx antennas, which are assumed to be isotropic by the 3GPP models. Under these conditions, diffraction effects are favored in short-range distances compared to cases involving high-directive antennas. Note that real systems operating at sub-THz bands are expected to incorporate highly directive antennas in point-to-point communications to compensate for the significant attenuation of these frequencies. Regarding the frequency dependence, the 3GPP models also underestimate its impact. The corresponding loss exponent is clearly lower in absolute terms than those obtained for the fitted models, indicating that the effects of human blockages are more severe in the sub-THz band than in the mmWave~range.

\begin{table}[t]
    \centering
    \renewcommand{\arraystretch}{1.75}
    \caption{Numerical values obtained \linebreak after fitting measurements and models}
    \begin{tabular}{|c|c|c|c|}
        \multicolumn{1}{c}{\textbf{FI}} & \multicolumn{1}{c}{\textbf{CI}} & \multicolumn{1}{c}{\textbf{ABG}} & \multicolumn{1}{c}{\textbf{CIF}} \\
        \hline
        \multirow{2}{*}{$A$\:=\:16.0\:dB} & \multirow{2}{*}{$\phi$\:=\:12.0\:dB/m} & $\alpha$\:=\:12.1\:dB/m & $a$\:=\:12.1\:dB/m \\
        \multirow{2}{*}{$n$\:=\:$-$3.09} & \multirow{2}{*}{$m$\:=\:$-$3.32} & $\beta$\:=\:$-$5.2\:dB & $b$\:=\:0.0196 \\
        & & $\gamma$\:=\:$-$3.09 & $c$\:=\:$-$3.32 \\ \hline\hline
        \red{$\sigma$\:=\:9.65\:dB} & \green{$\sigma$\:=\:6.18\:dB} & \green{$\sigma$\:=\:6.17\:dB} & \green{$\sigma$\:=\:6.17\:dB} \\ \hline
        \multicolumn{4}{c}{\vspace{-1.5em}} \\
    \end{tabular}
    \begin{tabular}{cc|c|c}
        & \multicolumn{2}{c}{\textbf{3GPP-A and 3GPP-B {R}MSEs}} & \\ \cline{2-3}
        \multicolumn{1}{c|}{} & $\mathrm{{R}MSE}_{\rm A}$\:=\:8.89\:dB & $\mathrm{{R}MSE}_{\rm B}$\:=\:8.97\:dB & \multicolumn{1}{c}{} \\ \cline{2-3}
    \end{tabular}
    \label{tab:numerical_results}
\end{table}

Finally, all the results obtained from fitting and estimating the blockage gain are summarized in Table \ref{tab:numerical_results}. For our site-specific conditions, the frequency loss exponent is consistently above 30\:dB/decade, while the distance gain factor is approximately 12\:dB/m. A different measurement setup might introduce additional contributions to the LoS, which could modify these values. For instance, a reflection not blocked jointly with the LoS component could reduce the frequency loss exponent, since it would not be affected by diffraction losses. Additionally, directive antennas with significantly different HPBW might also affect our results. Regarding models accuracy, the highest standard deviation, $\sigma_{\rm FI}$\:=\:9.65\:dB, is obtained for the FI model, since it does not account for distance. When this factor is incorporated, $\sigma$ decreases by approximately 3.5\:dB. Conversely, models based on the geometry of the scenario, such as 3GPP-A and 3GPP-B, perform worse, yielding RMSEs around 8.9\:dB. This is approximately 2.7\:dB higher than the standard deviation by the frequency and distance dependent models (CI, ABG, and CIF).

\section{Conclusions}
The inclusion of human blockage effects in next-generation channel models is essential to accurately predict the wireless propagation behavior, particularly at the sub-THz frequencies. To the best of the authors' knowledge, this work has conducted for the first time a measurement campaign analyzing human blockage across an extremely wide band, from 75\:GHz to 215\:GHz, encompassing both mmWave and sub-THz frequencies. Results show that additional attenuation due to human presence at distances from 1\:m to 2.5\:m ranges between 42\:dB and 56\:dB, significantly affecting the received power. Several models (FI, CI, ABG and, CIF) were fitted to the experimental values of blockage gain, consistently yielding a frequency loss exponent of approximately $-$3 and distance gain factor of 12\:dB/m. These results were compared with the standardized models 3GPP-A and 3GPP-B (based on site-specific geometry), which exhibited root mean squared errors approximately 2.7\:dB higher. The findings of this work are expected to aid in the development of D2D, BAN, and PAN communications systems by incorporating blockage gain as an additional term in the link-budget evaluations. As future \blue{works}, aspects such as the antenna height or larger Tx-Rx distances could also be evaluated \blue{for their effect on blockage~gain}.

\end{document}